\documentclass{nature}

\usepackage{amsmath}
\usepackage[pdftex]{graphicx}
\usepackage{dcolumn}
\usepackage{bm}     
\usepackage{dsfont}
\usepackage{color}
\usepackage{braket}
\usepackage{hyperref}

\newcounter{defcounter}
\usepackage{amsthm}

\title{Intermolecular Vibrations Drive Ultrafast Singlet Fission}

\author{Hong-Guang Duan$^{1,2,3,*}$, Ajay Jha$^{1,*}$, Xin Li$^{4}$, Vandana Tiwari$^{1,5}$, Hanyang Ye$^{6}$, Pabitra K. Nayak$^{6}$, Xiao-Lei Zhu$^{4}$,  Zheng Li$^{1}$, Todd J. Martinez$^{4}$, Michael Thorwart$^{2,3}$ \& R. J. Dwayne Miller$^{1,3,7}$}  

\begin{document}

\maketitle

\begin{affiliations}
\item Max Planck Institute for the Structure and Dynamics of Matter, Luruper
Chaussee 149, 22761, Hamburg, Germany 
\item I.\ Institut f\"ur Theoretische Physik,  Universit\"at Hamburg,
Jungiusstra{\ss}e 9, 20355 Hamburg, Germany
\item The Hamburg Center for Ultrafast Imaging, Luruper Chaussee 149, 22761
Hamburg, Germany
\item Department of Chemistry and PULSE Institute, Stanford University, Stanford, California 94305, United States
\item Department of Chemistry, University of Hamburg, Martin-Luther-King Platz 6, 20146 Hamburg, Germany 
\item Clarendon Laboratory, Department of Physics, University of Oxford, Parks Road, Oxford OX1 3PU, United Kingdom
\item The Departments of Chemistry and Physics, University of Toronto, 80 St.
George Street, Toronto Canada M5S 3H6\\
$^*$These authors contributed equally to this work. \\ 
\centerline{\underline{\date{\bf \today}}} 
\end{affiliations}

\begin{abstract} 
Singlet fission is a spin-allowed exciton multiplication process in organic semiconductors that converts one spin-singlet exciton to two triplet excitons. It offers the potential to enhance solar energy conversion by circumventing the Shockley-Queisser limit on efficiency. Recently, the mechanism of the primary singlet fission process in pentacene and its derivatives have been extensively investigated, however, the nature of the primary ultrafast process in singlet fission is still a matter of debate. Here, we study the singlet fission process in a pentacene film by employing a combination of transient-grating (TG) and two-dimensional (2D) electronic spectroscopy complemented by quantum chemical and nonadiabatic dynamics calculations. The high sensitivity of heterodyne detected TG spectroscopy enabled us to capture the vibrational coherence and to show that it mediates the transition from the singlet excited electronic state to the triplet-pair state. This coherent process is further examined by 2D electronic spectroscopy. Detailed analysis of the experimental data reveals that significant vibronic couplings of a few key modes in the low- and high-frequency region connect the excited singlet and triplet-pair states. Based on quantum chemical calculations, we identify these key intermolecular rocking modes along the longitudinal molecular axis between the pentacene molecules. They play the essential role of an electronic bridge between the singlet and triplet-pair states. Along with high-frequency local vibrations acting as tuning modes, these rocking motions drive the ultrafast dynamics at the multidimensional conical intersection in the singlet fission process. 
\end{abstract} 

Conversion of sunlight into electrical current with photovoltaic cells is one of the promising technologies to harness renewable energy. Concerted research efforts are directed to obtain cost-effective third generation photovoltaics with improved power conversion efficiency. Singlet fission (SF)\cite{Chem_Rev_110_6891_(2010)} is an exciton multiplication process, which could be used to overcome the fundamental thermodynamic constraint on efficiency, known as the Shockley-Queisser limit ($\sim30\%$ for an ideal single junction silicon cell)\cite{JPCL_6_2367_(2015)}. The SF process has been widely observed in acenes\cite{Nat_Chem_5_1019_(2013), Nat_Chem_9_341_(2016)}, diphenylisobenzofurans\cite{ARPC_64_361_(2013)}, carotenoids\cite{JACS_137_5130_(2015)}, and other conjugated molecules\cite{Nat_Chem_8_1120_(2016), JACS_139_663_(2017), JPCL_6_360_(2015)}.  In the singlet fission process, a  photoexcited singlet exciton ($\rm S_{1}$) is converted to two triplet excitons ($\rm 2 \times T_{1}$). The generally accepted mechanism of SF involves two steps: (1) primary singlet fission (PSF), which is a spin-conserved step; and (2) spin decoherence (SD) processes, 
\begin{eqnarray}
\label{eq:reduced Hamiltonian}
\ket{S_{0}}\xrightarrow{h\nu} \ket{S_{1}}\xrightarrow{\rm PSF}\ket{^{1}(T_{1}T_{1})}\xrightarrow{\rm }\ket{^{1}(T_{1}...T_{1})}\xrightarrow{\rm SD}\ket{T_{1}}+\ket{T_{1}}.  
\end{eqnarray}
In PSF, the singlet excited state ($\rm S_{1}$) is converted into a doubly excited pair of a spin-correlated triplet-pair state $\rm ^{1}(T_{1}T_{1})$, which is often referred to as the multiexciton state. This spin-correlated triplet-pair $\rm ^{1}(T_{1}T_{1})$ with an overall singlet spin acts as an intermediate to an interacting triplet-pair $\ket{^{1}(T_{1}...T_{1})}$ and finally leading to the generation of free triplet excitons ($\rm T_{1}+T_{1}$) via spin decoherence\cite{Nat_Phys_13_176_(2017), Nat_Phys_13_182_(2017), JACS_139_11745_(2017), JPCL_7_2370_(2016)}. If the energy levels in the system are such that $\rm 2E(T_{1})<E(S_{1})$, SF is an overall exothermic process. On the other hand, if $\rm 2E(T_{1})>E(S_{1})$, the singlet fission process is endothermic. The detailed molecular insight of the process holds the key to the realization of synthetic control over the process. Thus, it becomes imperative to reveal the underlying mechanisms of the singlet fission processes in different molecular systems. 

Pentacene and its derivatives have shown great promise as potential SF based photovoltaics. Baldo and co-workers have demonstrated an external quantum efficiency of $>126\%$ based on pentacene/$\rm C_{60}$ in photovoltaics\cite {Science_340_334_(2013), APL_103_263302_(2013)}. This level of effective quantum efficiency for charge collection has fueled the great interest in pentacene and its derivatives in determining the underlying photophysical mechanisms for the SF process\cite{ACR_46_1330_(2013)}. In crystals of pentacene and its derivatives, SF is an exothermic process, i.e., $\rm 2E(T_{1})<E(S_{1})$. This makes SF in pentacene energetically favorable and the process of PSF occurs on the time scale of 100 fs to form the spin-correlated triplet pair $\rm ^{1}(T_{1}T_{1})$ \cite {CPL_241_84_(1995), ACR_46_1330_(2013)}. This $\rm ^{1}(T_{1}T_{1})$ pair is shown to remain bound for hundreds of picoseconds, before spin decoherence forms $\rm 2 \times T_{1}$\cite {JACS_140_2326_(2018)}. There have been numerous studies to unravel the dynamics of the $\rm ^{1}(T_{1}T_{1})$ formation. In 2012, Zhu and co-workers have reported the direct observation of the $\ket{\rm ^{1}(T_{1}T_{1})}$ state using time-resolved two-photon photoemission spectroscopy \cite{Nat_Chem_4_840_(2012), ACR_46_1321_(2013)}. They have measured $\sim20$ fs rise time in the triplet population in both tetracene and pentacene, which they attributed to the formation of the multiexciton state. Based on this, they proposed a coherent mechanism in which the initially photogenerated exciton is converted to the triplet pair by a coherent superposition of the singlet and multiexciton states \cite{JCP_138_114102_(2013), JCP_138_114103_(2013), JCP_141_074705_(2014)}. However, first principles calculations suggested that, instead of the superexchange between singlet and triplet pair, strong electronic couplings to charge transfer states could possibly mediate the evolution of the singlet exciton to the multiexciton state \cite{JACS_136_5755_(2014)}. Moreover, a phenomenological model has been constructed to reproduce the time scale of SF in TIPS-pentacene \cite{PRL_115_107401_(2015)}. These results support a scenario where SF is mediated by a coherent superexchange mechanism via an energetically higher-lying charge transfer state. In addition, the role of the charge transfer state in the PSF has been further advocated by studies on intramolecular singlet fission in bipentacene molecules \cite{PNAS_112_5325_(2015), JACS_137_8965_(2015), Nat_Comm_7_13622_(2016)} and pentacene crystals \cite{Chem_Sci_9_1242_(2018)}. 

In contrast to these ideas, recently, vibronic coherence has been considered as a mediator between the singlet and triplet-pair states. Based on the vibrational analysis, Kukura and co-workers have proposed that the ultrafast SF in pentacene and its derivatives is realized by a conical intersection between the singlet exciton and triplet-pair state \cite{Nat_Phys_11_352_(2015)}. In this work, however, the details of vibrational dynamics was absent. Moreover, 2D electronic spectroscopy has been applied to study the SF of a pentacene crystal \cite{Nat_Chem_8_16_(2016)}. In this study, intramolecular ground-state vibrations are identified and the associated vibronic coherence has been considered to mediate the PSF from the singlet to triplet-pair states within a single molecule. Based on this, the vibronic dimer model has been constructed on the basis of intramolecular vibrations to fit the time scale of PSF. In addition, this vibronic coherence has also been reported in tetracene \cite{Nat_Chem_9_1205_(2017)} and hexacene \cite{Nat_Chem_9_341_(2017)}. The dynamics of singlet fission in rubrene has been examined by transient absorption spectroscopy, the results demonstrated that the PSF is mediated by a symmetry-breaking mode \cite{Nat_Chem_9_983_(2017)}. Apparently, despite these enormous theoretical and experimental efforts, the understanding of the primary step of SF, i.e., the dynamics of the formation of the $\rm ^{1}(T_{1}T_{1})$ state and the role of intermolecular vibrations in pentacene is still unclear. 

To clearly distinguish the different mechanistic aspects of ultrafast singlet fission in pentacene, we examine the process of PSF in a pentacene film using a combination of ultrafast heterodyne-detected TG and 2D electronic spectroscopy. Due to the near-zero background and associated higher sensitivity relative to pump-probe methods \cite{Nat_Chem_7_980_(2015)}, TG spectroscopy allows us to directly capture the small signal modulations due to the nuclear vibrational coherence that drives the singlet exciton to the triplet-pair state. Based on a vibrational analysis, several key modes are identified in the low- and high-frequency region, which mediate SF from the singlet exciton to the triplet-pair state. Moreover, the SF process is further examined by 2D electronic spectroscopy and the associated correlation analysis reveals the existence of vibronic couplings of key modes connecting the singlet exciton and triplet-pair state. Quantum chemistry calculations reveal the significant strength of the vibronic coupling of the molecular rocking vibrations along the longitudinal molecular axis in a pentacene dimer, which dramatically modulates the electronic coupling between the singlet exciton and triplet-pair state and works as the intermolecular vibration. With the assistance of vibronic couplings of intramolecular vibrational modes, intermolecular vibrations open up  a new channel, which drives SF in pentacene on an ultrafast timescale. 

\section*{Results}
\begin{figure}[h!]
\begin{center}
\includegraphics[width=14.0cm]{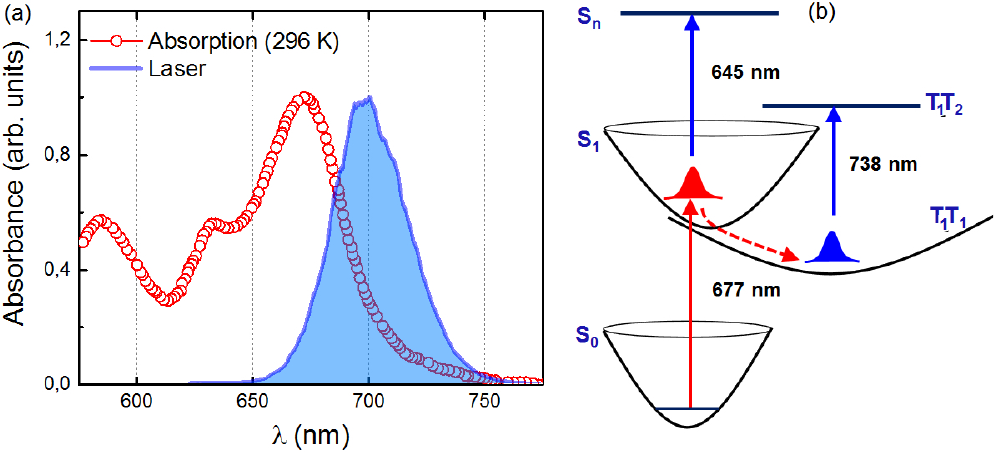}
\caption{\label{fig:Fig1} (a) Ground-state absorption spectrum of a pentacene film on a quartz substrate (red line) and laser spectrum used for the measurement (light-blue shadow). (b) Calculated site energies of ground, singlet excited states ($S_{0}$, $S_{1}$ and $S_{n}$) and triplet-pair states ($^{1}(T_{1}T_{1})$ and $^{1}(T_{1}T_{2})$). }
\end{center}
\end{figure}

Thin films of pentacene were prepared over 2 mm thick quartz substrates using the thermal evaporation method. The details of the sample preparation and their characterization are given in the Materials and Methods section. Fig.\ \ref{fig:Fig1}(a) represents the ground-state absorption spectrum (red circles) of a 200 nm thin pentacene film at room temperature (296 K). The lowest singlet state for pentacene lies at 677 nm ($\sim 14,800$ cm$^{-1}$). To perform transient spectroscopic measurements on these pentacene thin films, a broadband pulse is generated by a home-built NOPA and then, judiciously tuned in frequency. The spatial and temporal chirp of the pulse is compensated by the combination of a prism pair and a deformable mirror. The laser pulse spectrum used for the measurements is shown as the blue shaded region in Fig.\ \ref{fig:Fig1}(a). Within the probe window (650 - 750 nm), the primary step of SF is monitored by capturing the ground-state bleach (GSB, $\rm S_{0} \rightarrow S_{1}$), stimulated emission and excited state absorption (ESA, $\rm ^{1}(T_{1}T_{1}) \rightarrow ^{1}(T_{1}T_{2})$) features, which are illustrated in Fig.\ \ref{fig:Fig1}(b). The absorption signatures of $\rm S_{1}\rightarrow S_{n}$ lies at $\lambda<645$ nm \cite{Nat_Chem_8_16_(2016)}, which is out of our probing range. 

\subsection{Transient-grating spectroscopy.} 
\begin{figure}[h!]
\begin{center}
\includegraphics[width=10.0cm]{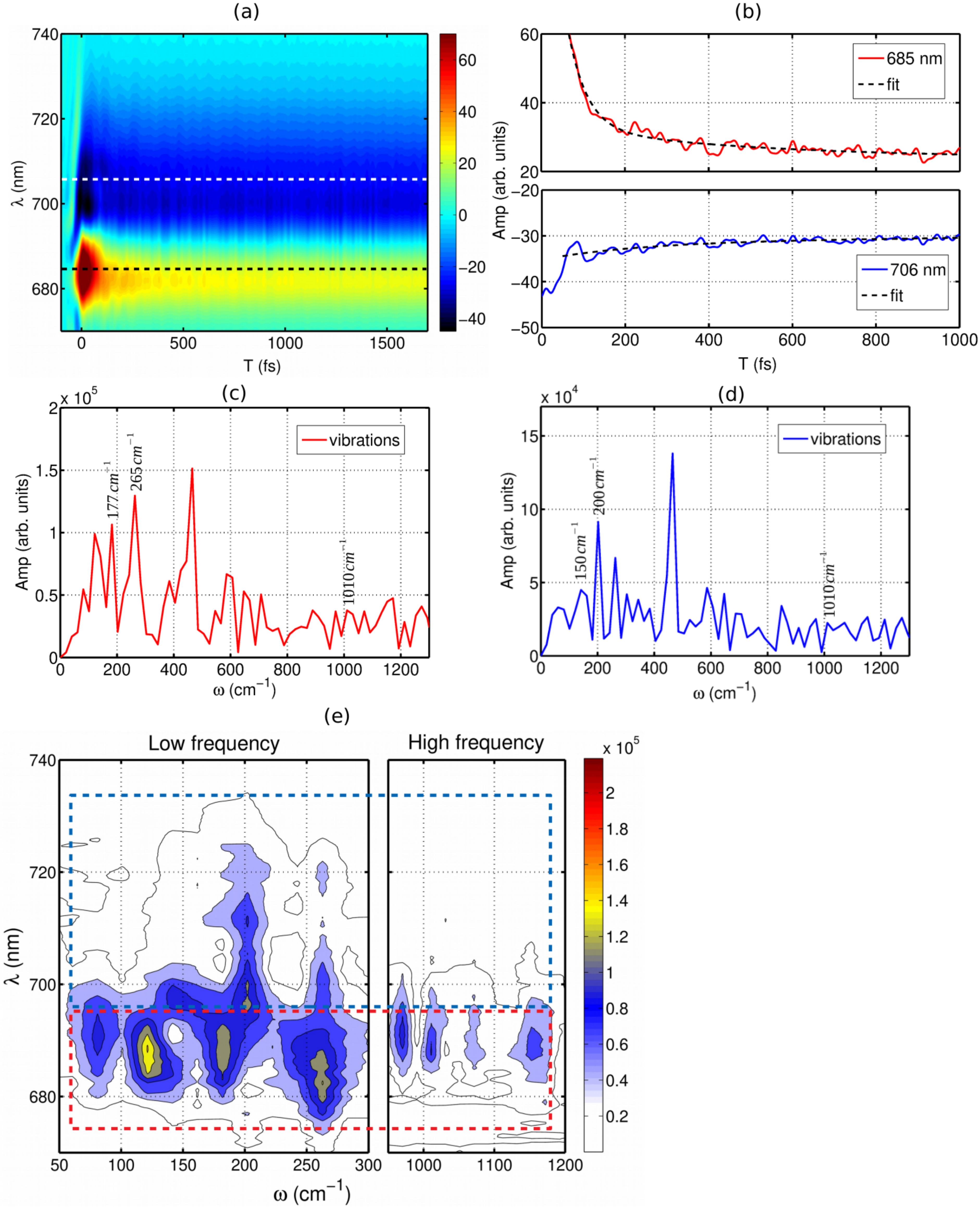}
\caption{\label{fig:Fig2} (a) Measured transient grating spectrum of the pentacene film with a time step of 2 fs. The ground state bleach (positive) and the excited state absorption (negative) in the spectrum originate from the corresponding transitions in Fig.\ \ref{fig:Fig1}(b) as red ($S_{0}\rightarrow S_{1}$) and blue arrows ($^{1}(T_{1}T_{1})\rightarrow ^{1}(T_{1}T_{2})$). (b) The selected kinetics of ground state bleach and excited state absorption at 685 and 706 nm, respectively. The associated traces are marked in the TG spectrum in (a). The vibrational frequencies of ground state absorption (685 nm) and excited state absorption (706 nm) are shown in (c) and (d), respectively, which are obtained by Fourier transform of residuals after removing the kinetics by the global fitting approach. (e) 2D vibrational map, obtained after a Fourier transform of the residuals along the probing window. The ground state bleach and excited state absorption areas are marked by red and blue boxes, respectively. }
\end{center}
\end{figure}

We measure the TG response of the pentacene thin film to capture the wave-packet dynamics associated with the transition of the singlet exciton $\rm S_{1}$ to the triplet-pair state $\rm ^{1}(T_{1}T_{1})$, also called the PSF. The details of the experimental setup and the measurement conditions are described in the Methods section. Fig.\ \ref{fig:Fig2}(a) shows the TG spectrum with positive (centered at 685 nm, $\sim14,600$ cm$^{-1}$) and negative bands (centered at 706 nm, $\sim14,180$ cm$^{-1}$). The TG spectrum evolves considerably in the initial 200 fs and then is largely invariant for the remaining times. The prominent rapid decay of the signal at $\sim14,600$ cm$^{-1}$ has been assigned in earlier reports to the decay of the stimulated emission from the singlet exciton due to PSF\cite{ACR_46_1330_(2013)}. The residual signal at $\sim14,600$ cm$^{-1}$ at later times corresponds to GSB. The ESA feature centered at $\sim14,180$ cm$^{-1}$ corresponds to the transition $\rm ^{1}(T_{1}T_{1}) \rightarrow ^{1}(T_{1}T_{2})$. Due to the enhanced sensitivity of TG measurements to capture coherences, the oscillations riding over these transient features are clearly visible in Fig.\ \ref{fig:Fig2}(a). Single-frequency traces of the GSB and ESA regions are shown in Fig.\ \ref{fig:Fig2}(b). To retrieve the coherent vibrational modes associated with the transient dynamics, we employ the global fitting approach on the TG spectrum and then, use Fourier transform analysis of the residuals after removing the decay components at all probe frequencies (for further description of the global fitting analysis, see the Supplementary Information). To show the excellent quality of the fits, two kinetic traces at $\lambda$ = 685 and 706 nm are shown along with the fitted decay curves in Fig.\ \ref{fig:Fig2}(b). The retrieved vibrational frequencies of GSB and ESA are shown in Fig.\ \ref{fig:Fig2}(c) and (d), respectively. The identified low- and high-frequency modes are marked accordingly. Moreover, the retrieved 2D power spectrum of the vibrations is shown in Fig.\ \ref{fig:Fig2}(e) along the probing window. This 2D power spectrum provides information on the existence of nuclear coherence in different regions of the probe frequencies (for a complete 2D vibrational map, see the Supplementary Information). In Fig.\ \ref{fig:Fig2}(e), the GSB and ESA bands are marked with red and blue dashed line boxes. The dominant vibrational modes observed in the GSB region are: 80, 121, 177, 262, 950 and 1010 cm$^{-1}$. Compared to the highly intense vibrational signals in the GSB, the fewer number of modes are seen in the ESA region. For instance, the 262 cm$^{-1}$ mode shows much less amplitude in the ESA band compared to the one in the GSB. This difference indicates that the mode is strongly damped after the wave-packet passes through the potential energy surface (PES) of the singlet exciton state to the triplet-pair state. Interestingly, two low-frequency modes at 120 and 177 cm$^{-1}$ show strong amplitudes in the GSB band and are absent from the ESA band. In addition, two new vibrational frequencies at 150 and 200 cm$^{-1}$ are observed in the ESA region, which suggest changes in the vibrational frequencies after the passage through the PESs from the singlet exciton state to the triplet-pair state. Thus, the 2D power spectrum retrieved from the TG response of pentacene thin films clearly captures the nuclear coherence in which the modes mediating the ultrafast processes of PSF can be identified by the relative degree of damping in the ground and excited state surfaces. 

\subsection{Two-dimensional electronic spectroscopy (2DES).} 
\begin{figure}[h!]
\begin{center}
\includegraphics[width=12.0cm]{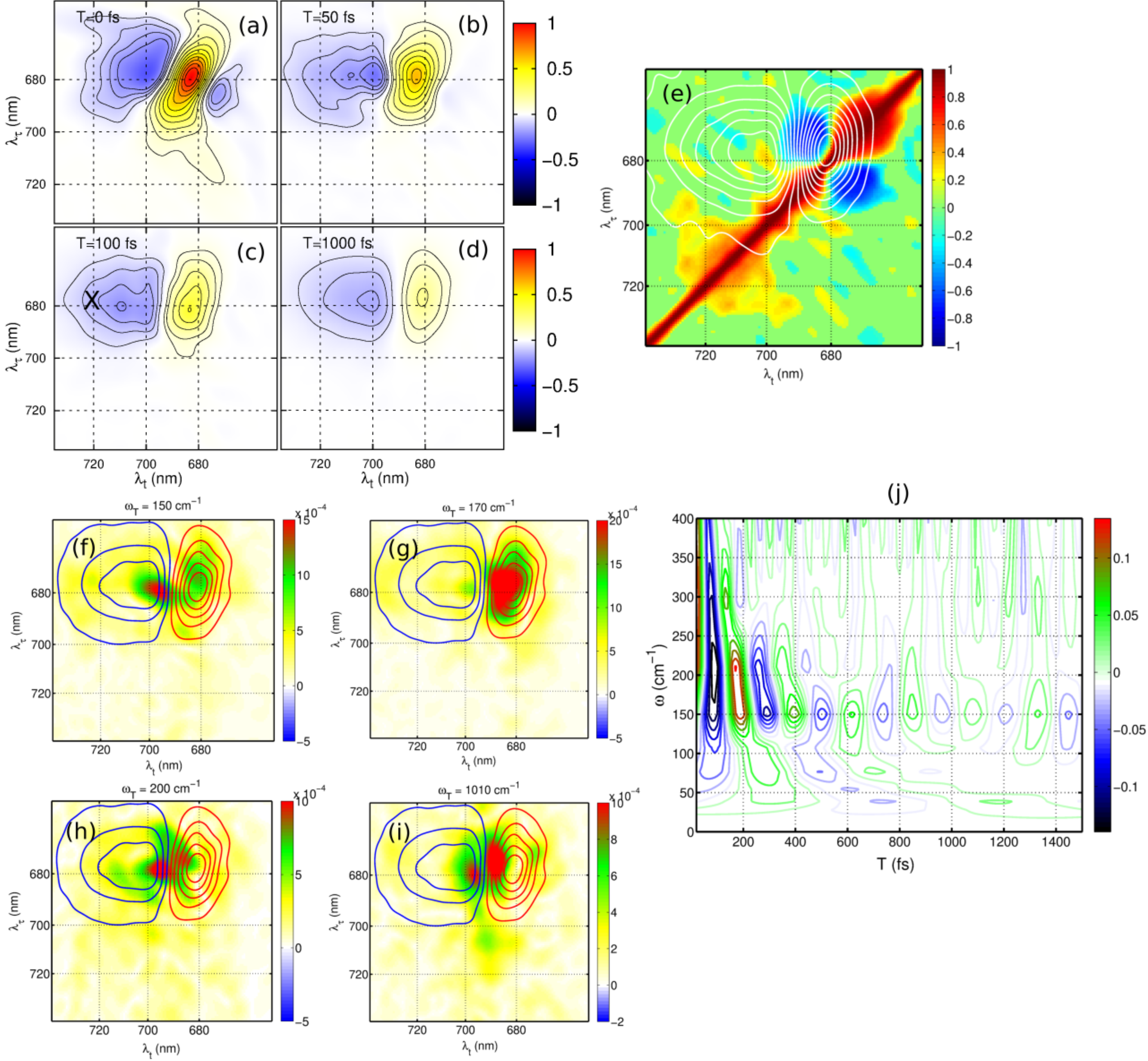}
\caption{\label{fig:Fig3} The selected 2D electronic spectra (real part) for selected waiting times are shown from (a) to (d). Positive diagonal and negative off-diagonal features denote the ground state bleach and excited state absorption, respectively. (e) 2D correlation map obtained from a correlation analysis along the diagonal direction. Two negative peaks are shown in the region of the ground state bleach (the 2D spectrum at $T = 500$ fs is shown as a white contour), which indicates the vibrational origin of the oscillations. The identified 2D vibrational maps of low frequencies, 150, 170, 200 cm$^{-1}$ and high frequency mode of 1010 cm$^{-1}$ are presented from (f) to (i), respectively. (j) Wavelet analysis of the trace at ESA (peak 'X' in (c)), ($\lambda_{\tau}$, $\lambda_{t}$) = (677, 722) nm in the 2D spectra. Instead of 170 cm$^{-1}$, two new frequencies are generated at 200 and 150 cm$^{-1}$, respectively. } 
\end{center}
\end{figure}

To further examine the coherent nuclear dynamics coupled with the primary electronic transition of SF, we measure the 2D electronic spectra of the pentacene film at room temperature (296 K) under similar experimental conditions as for the TG spectroscopy (for details on experimental conditions, see the Methods section). 2DES spectroscopy provides additional resolution along the excitation frequency axis, which helps to decongest the spectra due to different overlapping electronic transitions. In Fig.\ \ref{fig:Fig3}(a), the real parts of the 2D electronic spectra are shown for selected waiting times. The GSB and ESA bands in the TG spectrum are now separated along the excitation ($\omega_{\tau}$) and probe ($\omega_{t}$) frequencies. At $T=0$ fs, the center peak at $680$ nm is strongly stretched along the diagonal direction, which illustrates the significant inhomogeneous broadening of the excited electronic transition in the pentacene film. The 2D spectrum at $T=0$ fs also carries a small contribution originating from the quartz substrate \cite{ACS_Phot_5_852_(2018)}. With increasing waiting times, the magnitude of the central diagonal peak is strongly decreased due to the decay of the stimulated-emission component (see the 2D spectrum at $T=50$ fs  in Fig.\ \ref{fig:Fig3}(b)). In addition, this decay is also partially due to the population transfer from the singlet state to the triplet-pair state. The ESA band in the 2D spectrum at $T=50$ fs, appearing as an off-diagonal feature at ($\lambda_{\tau}$, $\lambda_{t}$) = (680, 700) nm, corresponds to the $\rm ^{1}(T_{1}T_{1}) \rightarrow ^{1}(T_{1}T_{2})$ transition. These ESA features allow us to follow the coherent generation of the $\rm ^{1}(T_{1}T_{1})$ state and subsequent dynamics over time. To further resolve the rates of the underlying electronic transitions, several series of consecutive 2D spectra at different times have been analyzed by the global fitting approach. The obtained time scale of the kinetics and the decay-associated spectra are shown in Section III of the Supplementary Information. The fastest decay component reveals a timescale of $\sim$100 fs, which is in good agreement with  earlier reports on the PSF process in pentacene \cite{Nat_Phys_11_352_(2015), Nat_Chem_8_16_(2016), PRB_84_195411_(2011)}. In addition, information on the coherent dynamics can be obtained by removing the kinetics from the measured data. Here, we first perform the 2D correlation analysis to distinguish the origin of coherence based on the correlation or anti-correlation of peaks along the diagonal direction (see the Supplementary Information for details) \cite{CP_545_40_(2012)}. In Fig.\ \ref{fig:Fig3}(e), the 2D correlation spectrum is shown together with the overlap of the 2D electronic spectrum at $T=500$ fs (white contour lines) to clarify the spectral position of the correlations. In the correlation map, two negative peaks (anti-correlation) with a strong magnitude are symmetrically located along the diagonal direction, which matches the peak of the GSB in the 2D spectrum at $\rm T = 500$ fs. Earlier theoretical work \cite{CP_545_40_(2012)} has demonstrated that the anti-correlation in the 2D correlation spectrum indicates the existence of vibrational coherence while the positive correlation corresponds to electronic coherence in the respective area of the 2D spectrum. In Fig.\ \ref{fig:Fig3}(e), there is no sign of any correlation in the position of the ESA peak, which signifies the absence of electronic quantum coherence mediating the generation of the triplet-pair state. Thus, our measurement does not support the conclusion of Ref.\ \cite{Nat_Chem_4_840_(2012)}. Furthermore, we perform the Fourier transform of the 2D residuals after removing the kinetics and obtain the 2D power spectra of different vibrational frequencies. A few of the dominating frequencies (150, 170, 200 and 1010 cm$^{-1}$) have been plotted as 2D vibrational maps in Fig.\ \ref{fig:Fig3}(f) to (i). The low-frequency mode at 170 cm$^{-1}$ (Fig.\ \ref{fig:Fig3}(g)) with a strong magnitude in the 2D power spectrum is located at the center peak of the GSB and does not show a similar magnitude at the ESA peak. It implies that this low-frequency mode cannot be transferred from the singlet excited state to the triplet-pair state unless the vibrational frequency has changed after the wave-packet reaching the triplet-pair state. Moreover, we observe the magnitude in 2D vibrational maps of 150 and 200 cm$^{-1}$ are mainly located in the region of ESA, which agrees to the observation from TG spectrum in Fig.\ \ref{fig:Fig2}(e). In addition, we observe the magnitude of vibration at 1010 cm$^{-1}$ is located both in GSB and ESA in Fig.\ \ref{fig:Fig3}(i). To further track the vibrational coherence, we carry out a wavelet analysis of the vibrational coherence on the ESA peak in the 2D spectra (see Supplementary Information for more details). We select the peak at ($\lambda_{\tau}$, $\lambda_{t}$) = (680, 722) (averaged in a square with $\pm 50$ cm$^{-1}$ and it  marked as `X' in Fig.\ \ref{fig:Fig3}(c)) and plot the result in Fig.\ \ref{fig:Fig3}(j). We observe that instead of 170 cm$^{-1}$, two new vibrational modes are generated at 200 cm$^{-1}$ and 150 cm$^{-1}$ after the wave-packet passes through the PES of the singlet excited state to the triplet pair state. In Fig.\ \ref{fig:Fig3}(j), the kinetics of the 200 cm$^{-1}$ mode reaches its maximum within 200 fs and then decays rapidly within 400 fs. More interestingly, the low-frequency mode at 150 cm$^{-1}$ reaches its maximum within 200 fs as well but persists for a longer lifetime (more than 1.5 ps). All the observations agree well with the TG measurement (the frequency resolution of the power spectra is slightly different in the TG and 2D electronic measurements, see Supplementary Information for more details). Moreover, as an example to describe the high-frequency modes, the 2D vibrational map of 1010 cm$^{-1}$ (Fig.\ \ref{fig:Fig3}(i)) shows a strong-magnitude link between the diagonal GSB and off-diagonal ESA feature, which manifests the population transfer of the vibrational coherence from the singlet excited state to the triplet-pair state. Thus, based on the 2D vibrational analysis of time series of 2D spectra, we clearly identify a few key vibrational modes in the low and high-frequency regime, which drive the coherent generation of the triplet-pair state. 

\subsection{Theoretical calculations and modeling.} 
\begin{figure}
\begin{center}
\includegraphics[width=10.0cm]{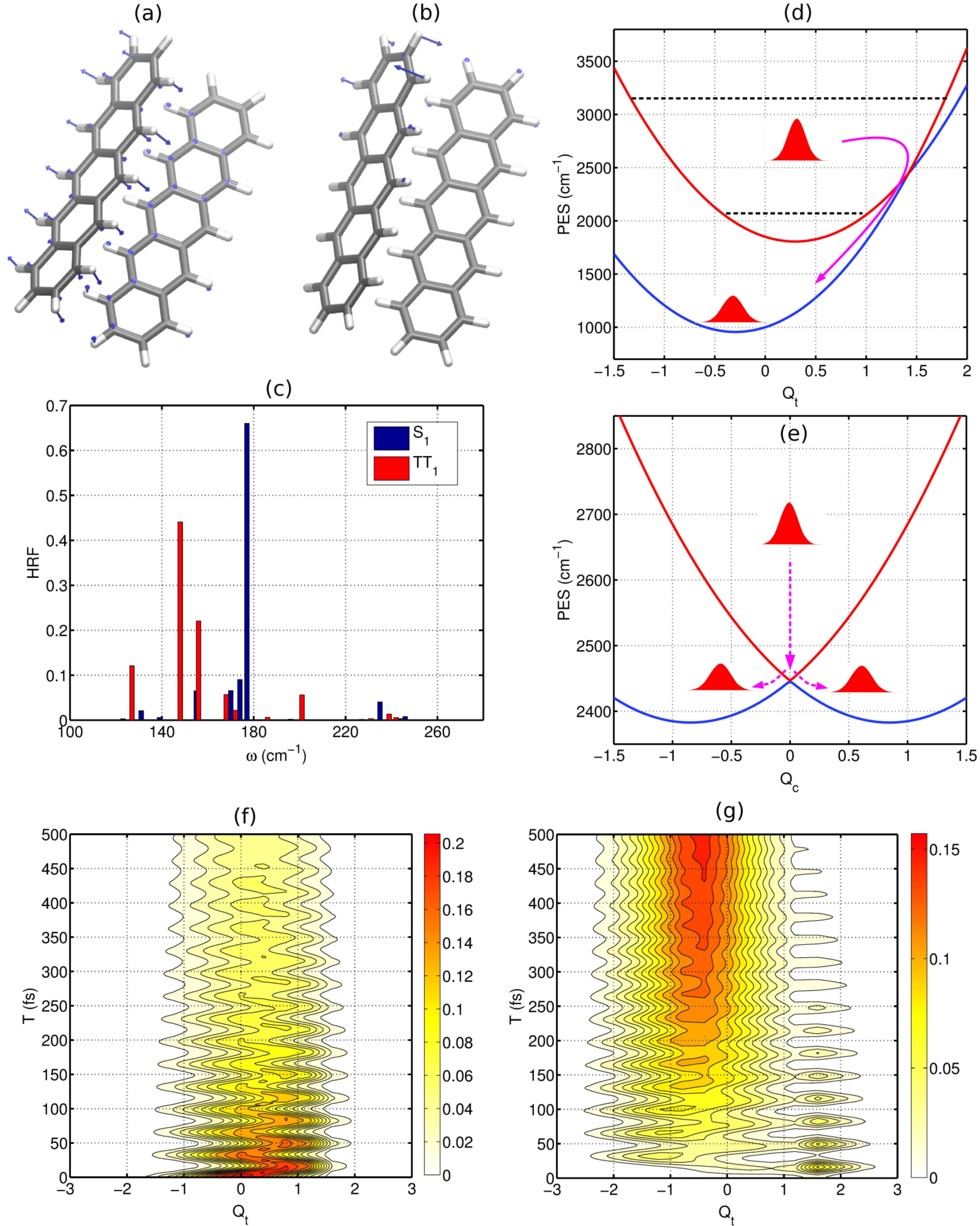}
\caption{\label{fig:Fig4} (a) and (b), few key modes identified from theoretical calculations. The low-frequency mode (177 cm$^{-1}$ in (a)) is associated with the intermolecular rocking motion of two pentacene molecules along the longitudinal molecular axis, which serves as the intermolecular vibration. The calculated high-frequency mode of 1013 cm$^{-1}$ (in (b)) corresponds to the intramolecular vibration. (c), the calculated Huang-Rhys factors of the low-frequency modes from 100 cm$^{-1}$ to 280 cm$^{-1}$. Blue and red bars correspond to the modes of the singlet and triplet-pair states. Constructed PES along the tuning (d) and the coupling (e) modes, which are based on quantum chemistry calculations for the Huang-Rhys factor between the singlet excited and triplet-pair states. The calculated wave-packet dynamics of $S_{1}$ and $^{1}(T_{1}T_{1})$ along the tuning mode are shown in (f) and (g), respectively. }
\end{center}
\end{figure}

To assign the observed vibrations both in the TG and the 2D electronic spectra, we have performed  quantum chemical calculations on a pentacene pair, see Fig.\ \ref{fig:Fig4}(a) and (b). The initial geometry of the pentacene pair has been extracted from the crystal structure\cite{JACS_129_10316_(2007)} and the ground state was optimized by a mixed quantum mechanical--molecular mechanical approach on the DFT level with general Amber force field for the pentacene dimer inside the crystal structure (See Figure S12 in the Supplementary Information). The calculated vibrational frequencies and the associated Huang-Rhys factors between the singlet and triplet-pair electronic states are shown in the Supplementary Information (see Section VIII). To account for the triplet-pair generation, an ab initio exciton model has been employed, which provides us with information on the molecular vibrations of the pentacene pair and the Huang-Rhys factors between the singlet excited state and triplet-pair state. Using these calculations, vibrational modes with strong Huang-Rhys factors are clearly identified, as shown in Fig.\ \ref{fig:Fig4}(c). Among them, the low-frequency mode of 177 cm$^{-1}$ corresponds to the intermolecular rocking motion in the pentacene dimer, which, based on our simulations, significantly modulates the electronic couplings between the singlet exciton and triplet-pair state (the details are  described in Section X of the Supplementary Information). Additionally, a few high-frequency modes (1013, 1196 and 1517 cm$^{-1}$) with strong Huang-Rhys factors are resolved as well (shown in the Supplementary Information). Based on the calculations, these high-frequency modes correspond to C--C and C=C stretching, which strongly modulate the site energies of the electronic excited state of the pentacene molecule and can be classified as intramolecular vibrations. These key modes have strong vibronic couplings of $\sim$100 meV to the electronic states (details are given in the Supplementary Information, Section VIII), which implies that such high-frequency modes can significantly reduce the effective energy gap between the singlet and triplet-pair states, and thus, make them even nearly degenerate in the vicinity of a conical intersection \cite{JPCL_6_4972_(2015)}. Moreover, to distinguish the vibrational coherence on each of the  $S_{1}$ and $^{1}(T_{1}T_{1})$ PES, we calculate the vibrational modes of both electronic states in Fig.\ \ref{fig:Fig4}(c). Interestingly, we find the vibrational mode of 177 cm$^{-1}$ with a strong Huang-Rhys factor.  This mode changes its frequency to 150 cm$^{-1}$ on the triplet-pair state PES (notice the correlation analysis of the vibrational modes in Fig.\ S8 of Supplementary Information), which is in excellent agreement with the observations revealed in our 2D spectroscopic measurements in Fig.\ \ref{fig:Fig3}(f) to (j). 

To illustrate the role of the identified key modes (inter- and intramolecular vibrations) both in theory and experiment, we construct a simple two-state two-mode model to calculate the wave-packet dynamics during the PSF process (more details of the calculation are described in the Section XI of the Supplementary Information). We construct the PESs with an energy difference of 850 cm$^{-1}$ and electronic coupling 6 cm$^{-1}$ \cite{ele_coupling} between the two electronic states. The reaction coordinates for the transition are selected according to the strong vibronic interaction of intramolecular vibrations with 1013 cm$^{-1}$ (although there are few more high-frequency modes) and the intermolecular vibration with 177 cm$^{-1}$. The corresponding intra- and intermolecular vibration can now be identified as tuning and coupling modes ($Q_{t}/Q_{c}$), respectively. Based on the computed vibronic coupling strengths, the conical intersection can be established at the reaction coordinate $ Q_{t}=1.4$, as shown in Fig.\ \ref{fig:Fig4}(d). In this figure, the wave packet has the possibility to pass through the PES of the singlet excited state to the triplet pair state when the vibrational coherence is generated between two vibrational levels of the tuning mode (marked by two dashed black lines). On the basis of these parameters, we simulate the coherent dynamics of the wave packet on the PESs. For this, we assume the initial wave packet to be on the lowest vibrational level of the singlet excited state and calculate the time evolution of the density matrix of the two-state two-mode model. Due to the strong vibronic couplings in the vicinity of the conical intersection, the wave-packet dynamics is calculated by the numerically exact hierarchy equation of motion approach \cite{JPSJ_74_3131_(2005)} using GPU parallelization \cite{JPCL_3_2828_(2012), JCTC_11_3859_(2015)}. The calculated wave-packet dynamics are shown in Fig.\ \ref{fig:Fig4}(f) and (g) on the singlet exciton state and the triplet-pair state, respectively. Based on the calculation, we observe that, initially, the wave packet is mainly located on the singlet excited state PES. With on-going time evolution, the wave packet propagates to the right side and reaches $\rm Q_{t}=1.4$ where it can decay to the lower excited state through the conical intersection. After that, the wave packet on the lower excited state moves further to the right side to $Q_{t}=2.5$ and then returns back to the degenerate point. The wave packet repeats the propagation process again to reach the CI and transition to the lower excited state. After 100 fs, we observe that the amplitude of the wave packet on the $^{1}(T_{1}T_{1})$ state gradually reaches its limit before 400 fs. The obtained timescale of the population transfer for the $\rm S_{1}$ to $\rm ^{1}(T_{1}T_{1})$ process is $\sim$180 fs (population dynamics of $S_{1}$ and $^{1}(T_{1}T_{1})$ are shown in the Supplementary Information), which qualitatively agrees with the SF timescale reported by other groups \cite{ACR_46_1330_(2013)}. Moreover, the motion of the wave packet is shown to have the period of $\sim$32 fs, which perfectly matches the vibrational frequency of the tuning mode in the modeling. Based on the calculations, we demonstrate the transfer of vibrational coherence of the intramolecular vibration from the singlet state to the triplet-pair state, which agrees with the observations in our 2D vibrational maps obtained from the TG and 2D electronic spectra. The wave-packet dynamics along the coupling mode is shown in the Supplementary Information (see Section XI). The results show that the vibrational coherence of the low-frequency coupling mode does not lead the PESs to transfer to the $\rm ^{1}(T_{1}T_{1})$ state, which is in complete agreement with our experimental observation from TG and 2D measurements (Fig.\ \ref{fig:Fig2}(e) and Fig.\ \ref{fig:Fig3}(f) to (i)). Thus, based on the kinetic analysis and coherence transfers, our model of the conical intersection fully reproduces the experimental observations revealed by TG and 2D electronic measurements. 

To further identify the functional role of the intermolecular vibrations, we turn off the vibronic coupling of the coupling mode by setting $\Lambda=0$ cm$^{-1}$ (the model parameters are described in the Supplementary Information). We find that the transfer of the wave-packet population is significantly slower than in the experiment ($\sim$100 fs) since the electronic coupling between the singlet excited state and triplet-pair state is weak. It underpins the fact that solely vibronic coupling to the electronic state is not sufficient to reproduce the ultrafast PSF dynamics\cite{Nat_Chem_9_1205_(2017)}. In addition, the theoretical model presented by Rao and co-workers\cite{Nat_Chem_8_16_(2016)} perfectly reproduces the timescale of PSF in pentacene, however, with rather strong electronic couplings between $\rm S_{1}$ and $\rm ^{1}(T_{1}T_{1})$. Such a scenario is not supported by the previous quantum chemistry calculations \cite{JACS_136_5755_(2014)}. Moreover, to monitor the role of the charge separation (CT) state, the population dynamics of PSF has been modeled by other groups. Their calculations show long-lived electronic quantum coherence between $\rm S_{1}$, $\rm CT$ and $\rm ^{1}(T_{1}T_{1})$ originating from the strong electronic couplings between $\rm CT$ and $\rm S_{1}$, $\rm ^{1}(T_{1}T_{1})$ \cite{PRL_115_107401_(2015)}. However, our correlation analysis of the measured 2D electronic spectra clearly demonstrates the absence of long-lived electronic coherence. In our study, the CT state does not show a significant participation in the PSF process. More important, except the intramolecular mode of 1013 cm$^{-1}$ used in our modeling, there are few more high-frequency modes that are identified both in the measurements of the TG and 2D electronic spectra.  With the intermolecular vibration at 177 cm$^{-1}$, these multi-vibrational modes generate a  conical intersection of hyper-dimensional PESs. Based on our work, a few high-frequency modes with strong vibronic couplings further reduce the barrier of the $S_{1}$ and $^{1}(T_{1}T_{1})$ states and the low-frequency modes bridge the two PESs to form the conical intersection, which drives SF on the ultrafast timescale\cite{multichannel}. 

We note here that our modeling is limited to a pentacene dimer. In pentacene films, the initially created exciton states are more delocalized, typically over a few (4-5) molecules, but the notion of a coherence length is ill-defined. Some estimate can be gained from the decay time of electronic quantum coherence observed in our 2D spectroscopic measurement. The antidiagonal linewidth in Fig.\ \ref{fig:Fig3}(a) at T=0 gives a homogeneous linewidth of $\sim170$ cm$^{-1}$ as lower estimate given the interference with excited state absorption\cite{PNAS_114_8493_(2017), Nat_Chem_6_196_2014}. This linewidth corresponds to the lifetime of electronic dephasing of $\sim 61$ fs, which is related to the coupling to the bath. The uncorrelated bath fluctuations limit the coherence length to a few pentacene molecules at any given instant. The absence of long-lived electronic coherence and subsequent SF demonstrate that the delocalized exciton rapidly collapses to a localized triplet-pair state during the SF process. Nevertheless, it would be interesting to extend the exciton basis beyond the model pentacene dimer to see what effect this would have on the required vibronic couplings. At present, such an extended basis is computationally beyond the scope of the present work. We expect that the basic physics would remain unchanged given the expected rapid decoherence and localization of the excitation. The present results illustrate that the ultrafast transition from singlet to triplet-pair state, mediated by a strong vibronic coupling of intermolecular vibration, destroys the electronic quantum coherence and spatial extent of the nascent exciton state \cite{Duan_submitted}. 

%

\section*{Conclusions} 
In summary, we have performed a detailed investigation of the mechanism of the primary step of singlet fission in a  pentacene film by using a combination of ultrafast heterodyne-detected TG and 2D electronic spectroscopy. The wave-packet dynamics has been monitored to capture the coherent $S_{1}\rightarrow ^{1}(T_{1}T_{1})$ crossing by tracking the vibrational coherence from the singlet exciton to spin-correlated triplet-pair state. Time-domain analysis of the vibrational transients reveal the existence of a novel low-frequency mode of 177 cm$^{-1}$ corresponding to the rocking motion along the longitudinal molecular axis in a pentacene dimer, with high frequency modes remaining unchanged. These observed key modes show strong vibronic couplings, which bridge the singlet exciton to triplet-pair state. In addition, based on theoretical calculations, the low-frequency rocking motion strongly modulates (tunes) the electronic coupling between the two electronic states. The high-frequency vibrational modes, which correspond to the intramolecular vibrations with significant vibronic couplings, reduce the energy gap between the singlet exciton and triplet-pair states and render them nearly degenerate. In conclusion, our work demonstrate in details (experimentally and  theoretically) that the concept of conical intersection can be constructed by the interplay of inter-molecular (low-frequency) and intra-molecular (high-frequency) modes coupled to the electronic DOFs, which dramatically speeds up the process of SF. Hence, our work emphasizes the importance of the low frequency modes, which can be used as an important design principle for developing new materials to achieve an efficient primary singlet fission process in new chemical architectures. 
%


\section*{Materials and Methods}
\subsection{Sample preparation.}
Quartz substrates were sonicated in a water solution of 2.5 $\%$ Hellmanex III soap, deionized water (DIW), acetone and isopropyl alcohol (IPA) each at 50$^\circ$C for 10 minutes. The cleaned substrates were blow-dried with compressed air and then transferred into a UV-Ozone cleaner (UVO-cleaner, Model 30, Jelight Company Inc.) for 10 minutes before being loaded into the vacuum chamber (base pressure approx. 5$\times$10$^{-6}$ mbar, “B30” from Oerlikon Leybold Vacuum Dresden GmbH, Germany, with major parts upgraded. This chamber opens up to air, i.e. samples fabricated in this chamber will see air when removed from the chamber for further measurements). Pentacene thin films of 200 nm thickness were obtained by thermal evaporation at the rate of 0.5 $\rm \AA$/s. A quartz crystal oscillator was used to monitor the evaporation rate. The evaporated thin films were transferred into a glove box with exposure to air as little as possible. Then the samples were kept sealed in N2 before they were used for time-resolved optical studies. XRD thin film measurements of pentacene were performed in Smartlab X-ray diffractometer (Rigaku Corporation) (the XRD data are shown in the Supplementary Information). The voltage and current of the instrument were set as 40 kV and 30 mA. X-ray wavelength is 1.54059 $\rm \AA$. UV-vis measurement of the samples was performed in a LAMBDA 1050 UV/Vis spectrophotometer (PerkinElmer Inc.). 

\subsection{Transient grating and 2D Electronic measurements with experimental conditions.} 
Details of the experimental setup have already been described in earlier reports from our group.\cite{PNAS_114_8493_(2017)} Briefly, the measurements have been performed on a diffractive optics based on an all-reflective 2D spectrometer with a phase stability of $\lambda/160$.\cite{UP_162_432_(2015)} The laser beam from a home-built nonlinear optical parametric amplifier (NOPA, pumped by a commercial femtosecond Pharos laser from Light Conversion) is compressed to $\sim$18 fs using the combination of a deformable mirror (OKO Technologies) and a prism pair. Frequency-resolved optical grating (FROG) measurement is used to characterize the temporal profile of the compressed beam and the obtained FROG traces are evaluated using a commercial program FROG3 (Femtosecond Technologies). A broadband spectrum so obtained carried a linewidth of $\sim$100 nm (FWHM) centered at 700nm which covered the electronic transitions to the first excited state and the excited state absorption in triplet-pair states. Three pulses are focused on the sample with the spot size of $\sim$100 $\mu$m and the photon echo signal is generated at the phase-matching direction. The photon-echo signals are collected using Sciencetech spectrometer model 9055 which is coupled to CCD linear array camera (Entwicklungsb{\"u}ro Stresing). The 2D spectra for each waiting time T were collected by scanning the delay time $ \tau = t_{1}-t_{2}$ in the range of [-128 fs, 128 fs] with a delay step of 1 fs. At each delay step, 100 spectra were averaged to reduce the noise ratio. The waiting time $ T = t_{3}-t_{2}$ was linearly scanned in the range of 2.1 ps with steps of 10 fs. For all measurements, the energy of the excitation pulse is attenuated to 10 nJ with 1 kHz repetition rates. Phasing of obtained 2D spectra was performed using an ``invariant theorem" \cite{JCP_115_6606_(2001)}. To optimize the contribution of ESA, the pentacene film has been slightly tilted (5$\sim$6$^{\circ}$) relative to the plane perpendicular to the incident beam. 

\subsection{Theoretical calculations.}  
The ab-initio exciton model \cite{JCTC_13_3493_(2017)} was used to study singlet fission in a pentacene crystal. In this approach, the elements of the exciton model Hamiltonian matrix (namely the excitation energies and couplings) are calculated in an ab-initio manner without empirical parameters from first principles. We have reported the inclusion of CT excited states in the exciton model, and in this work, the multiexcitonic excited states are further involved to properly describe the singlet fission process. The ab-initio exciton model provides a more efficient approach as compared to the multiconfigurational methods and minimizes the errors without introducing empirical parameters in the Hamiltonian matrix. 

We employed a QM/MM setup where a pentacene dimer is described by the ab-initio exciton model, and the surrounding pentacene molecules are modeled by the general Amber force field \cite{JCC_15_1157_(2004)} (for definition of the QM/MM regions, see the Supplementary Information). For the QM region, the long-range corrected (LRC) density functional LRC-$\omega$ PBEh \cite{JCP_130_054112_(2009)} and the cc-pVDZ basis set \cite{JCP_90_1007_(1989)} were used in density functional theory (DFT) and time-dependent (TD) DFT calculations, with dispersion correction modeled by Grimme's dispersion \cite{JCP_132_154104_(2010)} with Becke-Johnson damping \cite{JCC_32_1456_(2011)}. Ab-initio exciton model calculations were carried out using the GPU-accelerated TeraChem code \cite{JCTC_5_1004_(2009)}. The exciton model for the pentacene dimer in the QM region was setup to involve two locally excited (LE) states (one on each pentacene molecule), two CT excited states and one TT excited state. We have documented the exciton model in details in our previous work \cite{JCTC_13_3493_(2017), Acc_Chem_Res_47_2857_(2014)}. The spin-adapted wave function of the TT state is employed to facilitate evaluation of the couplings. Diagonalization of the exciton model Hamiltonian gives the adiabatic excited states as linear combinations of the diabatic excited state (namely the LE, CT and TT states). The gradient of the ab-initio exciton model was derived based on the Hellmann-Feynman theorem. 

Comparison with the multiconfigurational method suggests that the ab-initio exciton model systematically overestimates the vertical excitation energies, while the ordering of the states and the energy gaps are nicely predicted by the model. In particular, the exciton model gives a Davydov splitting of 0.12 eV, which corresponds well with the experimental measurement and the multiconfigurational result \cite{JACS_136_5755_(2014)}. We have optimized the geometries and performed a numerical frequency analysis for the adiabatic S1 and S2 states of the QM pentacene dimer, which are of dominant TT and LE character, respectively. The Huang-Rhys factors $S_{i}$ were calculated as $S_{i} = m_{i}\omega_{i}\Delta Q_{i}^{2}/2\hbar$, where $m_{i}$, $\omega_{i}$ and $\Delta Q_{i}$ are the reduced mass, frequency and the equilibrium displacement of the i-th vibrational mode, respectively. For the calculation of the coherent dynamics, the population dynamics of the wave-packet in the two-state two-mode model are calculated by the hierarchical equation of motion \cite{JPSJ_74_3131_(2005)} with GPU parallelization (Tesla K80). To obtain the stable Hamiltonian matrix, 20 and 10 vibrational levels are used in the tuning and coupling ($Q_{t}/Q_{c}$) modes. To monitor the wave-packet dynamics on two reaction coordinates, we perform the projection calculation based on the theoretical work of Ref.\ \cite{JCP_93_1628_(1990), JCP_147_074101_(2017)}. 

\subsection{Data availability} All relevant experimental data and code for theoretical calculations are available from the authors upon request.

%
\begin{addendum}
 \item This work was supported by the Max Planck Society and the Excellence Cluster ``The Hamburg Center for Ultrafast Imaging - Structure, Dynamics and Control of Matter at the Atomic Scale" of the Deutsche Forschungsgemeinschaft. H-G. D. acknowledges financial support by the Joachim-Hertz-Stiftung Hamburg within a PIER fellowship. The authors thank V. I. Prokhorenko for help with the 2D setup and for providing the 2D data analysis software. The authors thank Prof. M. Riede for help in the sample preparation. 

\item[Supporting information] The Supplementary Information includes the X-ray diffraction data  of the pentacene film, the description of the global fitting approach and the fitted kinetic traces, calculated vibrational modes and its Huang-Rhys factors and the modeling of the  wave-packet dynamics on the PESs.

\item[Competing Interests] The authors declare that they have no competing financial interests.

\item[Correspondence] Correspondence of theoretical part should be addressed to M.T. ~(email:michael.thorwart@physik.uni-hamburg.de) and Z.L. ~(email:zheng.li@mpsd.mpg.de), Correspondence and requests for experimental details should be addressed to  R.J.D.M.~(email: dwayne.miller@mpsd.mpg.de) 
\end{addendum}
%

\newpage

\newpage


\begin{thebibliography}{11}
\bibitem {multichannel} Based on our two-state two-mode model, the transfer with time $\sim$180 fs of the PSF is slightly slower than the observation of $\sim$100 fs. With few high-frequency modes, we expect that a multi-channel modeling of a conical intersection speeds up the population transfer. However, the computational costs of the simulations are too large with  the numerically exact method.
\end{thebibliography}
\end{document}